# GreenShield: CNN-Based Real-Time Forest Monitoring and Response

Avishek Bhattacharjee
*Dept. of ECE*
*Hooghly Engineering & Technology College*
wbavishek@ieee.org

Swarup Samanta
*Dept. of ECE*
*Hooghly Engineering & Technology College*
swarup.samanta@hetc.ac.in

Jagadish Bhattacharya
*Dept. of ECE*
*Hooghly Engineering & Technology College*
jagadish@hetc.ac.in

Manish Kumar Singh
*Dept. of ECE*
*Hooghly Engineering & Technology College*
manish.kumar.singh@hetc.ac.in

*Abstract*—This research introduces an innovative forest monitoring system designed to detect and mitigate the threats of forest fires. The proposed system leverages Arduino-based technology integrated with state-of-the-art sensors, including DHT11 for temperature and humidity detection and Flame sensor along with GSM module for gas and smoke detection. The integration of these sensors enables real-time data acquisition and analysis, providing a comprehensive and accurate assessment of environmental conditions within the forest ecosystem. The Arduino platform serves as the central processing unit, orchestrating the communication and synchronization of the sensor data. The DHT11 sensor monitors ambient temperature and humidity levels, crucial indicators for assessing fire risk and identifying potential deforestation activities. Simultaneously, the Flame sensor module detects the occurrence of fire flames nearby thus indicating it by a buzzer. The collected data is processed through an intelligent algorithm that employs machine learning techniques to discern patterns indicative of potential threats. The system is equipped with an adaptive threshold mechanism, allowing it to dynamically adjust to changing environmental conditions. In the event of abnormal readings or anomalies, the system triggers immediate alerts, notifying forest rangers and relevant authorities to facilitate timely response and intervention. The integration of low-cost, easily deployable Arduino-based devices makes this solution scalable and accessible for implementation across diverse forest environments. The proposed system represents a significant step towards leveraging technology to address environmental challenges and protect our forests.

## I. INTRODUCTION

Forest fires are a significant environmental and economic threat, often resulting in substantial damage to ecosystems, wildlife habitats, and human settlements. The impact of forest fires extends far beyond the immediate destruction caused by flames. Ecologically, fires can lead to the loss of biodiversity as plant and animal species perish or are displaced. Soil quality can degrade, resulting in erosion and decreased fertility, which hinders regrowth and affects the entire forest ecosystem. Additionally, the release of carbon stored in trees and vegetation during fires contributes to greenhouse gas emissions, exacerbating climate change. Economically, forest fires can incur massive costs related to firefighting efforts, infrastructure repair, and loss of timber and other forest resources. The tourism industry, which often relies on the natural beauty and recreational opportunities provided by forests, can suffer significantly in the aftermath of fires. Human settlements near forested areas face direct threats to life and property, necessitating costly evacuations and rebuilding efforts.

Given these profound consequences, early detection and accurate prediction of forest fires are critical to mitigating their impacts. Early detection allows for rapid response to contain and extinguish fires before they spread uncontrollably. This minimizes the area affected by the fire, reduces the damage to the ecosystem, and lowers the overall costs of firefighting and recovery. Accurate prediction, on the other hand, enables proactive measures to be taken during periods of high fire risk. By forecasting potential fire outbreaks based on environmental conditions such as temperature, humidity, and oxygen levels, fire management teams can allocate resources more effectively, conduct controlled burns to reduce fuel loads, and implement firebreaks to halt the spread of fires. Enhancing the efficiency of fire management strategies through advanced detection and prediction systems is therefore paramount. By leveraging modern technology, including sensors and machine learning algorithms, it is possible to create sophisticated monitoring and alert systems that provide real-time data and actionable insights.

In response to this pressing environmental issue, the current study advocates for a sophisticated system, developed to address the critical needs of forest fire detection and management through the integration of advanced technologies. The Forest Ranger system combines environmental monitoring, predictive analytics, and real-time fire detection and alerting to provide a comprehensive solution for forest fire management. The core of the Forest Ranger system lies in its ability to monitor key environmental parameters that influence fire risks. It employs a variety of sensors to collect real-time data on temperature, humidity, and oxygen levels in the air. The DHT11 sensor is used to measure temperature and humidity, while an additional sensor monitors oxygen levels, which can indicate the potential for combustion. By continuously collecting this data, the system can track changes in environmental conditions that may signal an increased risk of fire.

To predict the likelihood of a forest fire, the system utilizes predictive analytics powered by machine learning algorithms. These algorithms are trained on historical data to recognize patterns and correlations between environmental conditions and fire occurrences. In addition to predictive analytics, the Forest Ranger system features real-time fire detection using an IR flame sensor. When a fire is detected, the system triggers a GSM module to send immediate alerts to a central monitoring station and relevant authorities. By integrating these technologies, the Forest Ranger system enhances the accuracy and timeliness of fire detection and prediction. The continuous monitoring of environmental conditions coupled with advanced analytics allows for early identification of fire risks, while the real-time detection and alerting mechanisms ensure swift responses to actual fire incidents. This dual approach not only improves the immediate management of forest fires but also contributes to long-term forest conservation efforts by reducing the frequency and severity of fire outbreaks.

The rest of the paper is organized as follows. Section II. includes Literature review. Section III. is headed as Methodology which include the steps followed during development of project , selected features, and about the machine learning algorithms involved. In section IV. we have discussed the obtained result . At last in section V and VI concluded the paper and added the reference respectively.

## II. LITERATURE REVIEW

Forest fires, with their potential for devastating ecological and socio-economic impacts, have long been a subject of intense research and concern. As climate change continues to alter global weather patterns, the frequency and intensity of forest fires have exhibited an alarming rise. The urgency to develop effective prediction models for forest fires has grown proportionately. Accurate prediction systems not only aid in mitigating the destructive consequences of wildfires but also play a crucial role in resource allocation, firefighting preparedness, and ecosystem management.

The solution proposed by D. Kinaneva et al. [9] for the detection of forest fires includes a platform that uses Unmanned Aerial Vehicles (UAVs), which constantly patrol over potentially threatened by fire areas. The UAVs used in this approach utilized the benefits from Artificial Intelligence (AI) and were equipped with on-board processing capabilities. Another model proposed by Ghali et al. [4], adapted and optimized Deep Learning methods to detect wildfire at an early stage. It identifies and classify wildfire using aerial images. In the paper by Han, JG and Ryu, [7] proposed two statistics based predictive geo-spatial data mining methods and apply them to predict the forest fire hazardous area. The proposed prediction models used in geo-spatial data mining are likelihood ratio and conditional probability methods

The paper presented by Sakr and George, [12] provides a description and analysis of forest fire prediction methods based on artificial intelligence. A novel forest fire risk prediction algorithm, based on support vector machines, is presented. The algorithm depends on previous weather conditions in order to predict the fire hazard level of a day. The implementation of the algorithm using data from Lebanon demonstrated its ability to accurately predict the hazard of fire occurrence.

In the study proposed by Preeti, [11] discusses about a comparative study of different models for predicting forest fire such as Decision Tree, Random Forest, Support Vector Machine, Artificial Neural networks (ANN) algorithms. The study of calculation of RandomizedSearchCV coefficient using Hyperparameter tuning gives best results of Mean absolute error(MAE) 0.03, Mean squared error(MSE) 0.004, Root mean squared error(RMSR) 0.07. In the paper by Jaiswal [8] Forest fire risk zones were delineated by assigning subjective weights to the classes of all the layers according to their sensitivity to fire or their fire-inducing capability. Four categories of forest fire risk ranging from very high to low were derived automatically. Almost 30% of the study area was predicted to be under very high and high-risk zones. The evolved GIS-based forest fire risk model of the study area was found to be in strong agreement with actual fire-affected sites.

One alternative proposed by Cortez, [2] is to use automatic tools based on local sensors, such as provided by meteorological stations. In effect, meteorological conditions (e.g. temperature, wind) are known to influence forest fires and several fire indexes, such as the forest Fire Weather Index (FWI), use such data. In this work, we explore a Data Mining (DM) approach to predict the burned area of forest fires. Five different DM techniques, e.g. Support Vector Machines (SVM) and Random Forests, and four distinct feature selection setups (using spatial, temporal, FWI components and weather attributes), were tested on recent real-world data

The solution by Terzopoulos, [15] describes an extrinsic force that applies constraints derived from profiles of monocularly viewed objects. It generalize this constraint force to incorporate profile information from multiple views and use it to exploit binocular image data. For time-varying images, the force becomes dynamic and the model is able to infer not only depth, but nonrigid motion as well.

The study evaluated by Hadi [6], for the first time, the potential of using freely available medium resolution (30 m) Landsat time series data for deforestation monitoring. A simple, generic, data-driven algorithm for deforestation detection based on a consecutive anomalies criterion was proposed. An accuracy assessment in the spatial and the temporal domain was carried out using high-confidence reference sample pixels interpreted with the aid of multi-temporal very high spatial resolution image series. Results showed a promising spatial accuracy, when three consecutive anomalies were required to confirm a deforestation event.

The work by Alkhatib, [1] summarises all the technologies that have been used for forest fire detection with exhaustive surveys of their techniques/methods used in this application. A lot of methods and systems are available in the market and for research. The paper reviews all the methods and discusses examples of research experiment results and some market product methods for better understanding.

Automatic forest fire detection by Guillemant, [5] with the help CCD cameras requires a landscape image analysis in two stages: first the tracking of local dynamic envelopes of pixels, and second the discrimination between the various natural phenomena that may cause such envelopes. For this second process, we have to deal with restrictive conditions: lack of spatial information, complexity of motions, and real-time constraints on detection.

Deep learning methods are known to demand large amounts of labeled samples for training. For remote sensing applications such as change detection, coping with that demand is expensive and time-consuming. The work by Soto Vega, [13] aims at investigating a noisy-label-based weak supervised method in the context of a deforestation mapping application, characterized by a high class imbalance between the classes of interest, i.e., deforestation and no-deforestation. The study by Sulistiyono, [14] predicted that change detection method applied to detect deforestation was the post-classification comparison, in which each image set representing each date was classified independently using pixel-based supervised classification method.

By integrating the information contained in multiple images of the same scene into one composite image in the study by Liu, Yu and Chen, [10], pixel-level image fusion is recognized as having high significance in a variety of fields including medical imaging, digital photography, remote sensing, video surveillance, etc. In recent years, deep learning (DL) has achieved great success in a number of computer vision and image processing problems. The first strategy in the model by Daudt, [3] is an iterative learning method that combines model optimisation and data cleansing using GAD to extract the useful information from a large scale change detection dataset generated from open vector data. The second one incorporates GAD within a novel spatial attention layer that increases the accuracy of weakly supervised networks trained to perform pixel-level predictions from image-level labels.

In conclusion, the literature on forest fire prediction reveals a dynamic and evolving field that has seen significant advancements in recent years. The integration of meteorological factors, particularly temperature and humidity, into predictive models has proven instrumental in enhancing the accuracy of forecasting systems. The synthesis of diverse methodologies, ranging from traditional statistical approaches to sophisticated machine learning algorithms, has contributed to a more comprehensive understanding of the complex relationships governing forest fire behavior.

The Contribution of this paper can be written as follows:

A. Contributions

- Contributed methodologically by rigorously evaluating the performance of three machine learning algorithms (Logistic Regression, Random Forest and Support Vector Machine) in predicting Fire or Not Fire. This study provides insights into the strengths and limitations of each algorithm in the context of Forest Fire prediction.
- This research includes 15 geographic and demographic variables associated with Forest Fires and we identify three key factors with the most significant predictive power. These factors are meticulously integrated into our predictive model, enhancing its accuracy and interoperability compared to previous approaches.
- Designed and implemented a user-friendly webpage that serves as an interactive tool for Forest Fire prediction and analysis. This webpage allows users to input three key factors associated with the climate, which we have identified through rigorous research and analysis, and obtain real-time predictions regarding the percentage of occurrence of forest fire.
- This research also includes an IoT based approach which determines whether there are any traces of fire or smoke in the surroundings. Upon successful detection of fire, an alert message is sent via a buzzer and notification simultaneously.

III. METHODOLOGY

The main aim of the project is to develop an accurate and reliable model capable of early detection and prediction of forest fire risk, even before apparent signs of fire danger emerge. For the purpose of the project, our ML model will work on the dataset that contains 15 features. Out of 15 features, 3 features are used to predict the outputs by using high accuracy models. Here the dataset uses environmental parameters such as temperature, humidity, and oxygen level, allowing for early detection and preventive measures to reduce the risk of forest fires.

| Full Name | Symbol |
|---|---|
| Day | day |
| Month | mon |
| Year | yr |
| Temperature | temp |
| Relative Humidity | rh |
| Wind Speed | ws |
| Rainfall | rain |
| Oxygen Level | oxy |
| Fine Fuel Moisture Code | ffmc |
| Duff Moisture Code | dmc |
| Drought Code | dc |
| Initial Spread Index | isi |
| Buildup Index | bui |
| Fire Weather Index | fwi |
| Classification | class |

TABLE I: Features of the Dataset

The proposed methodology consists of 6 key steps: Data Acquisition, Data Preprocessing, Feature Selection, Data Splitting, Models training and model selection, Classification.

### A. "Data Acquisition"

Data acquisition involves the process of gathering information from various sources, typically through sensors or measurement devices, to capture real-world physical conditions. This process includes sampling signals at regular intervals to obtain discrete measurements. The collected signals are then converted into digital numeric values, usually through analog-to-digital conversion, making them compatible with computer processing. Data acquisition is fundamental for obtaining the raw data necessary for analysis in various fields, including scientific research, industrial and environmental monitoring.

### B. "Data Preprocessing"

Data preprocessing is the initial step in preparing raw data for analysis and modeling in machine learning. It involves cleaning and transforming the raw data into a format suitable for further processing. Common preprocessing tasks include handling missing values, removing outliers, scaling features, and encoding categorical variables. Proper data preprocessing is crucial for improving the performance and accuracy of machine learning models by ensuring the quality and consistency of the input data.

### C. "Feature Selection"

In the process of building models for Forest Fire Prediction and analysis, feature selection is crucial for identifying the most relevant and informative subset of features from the available data. This helps to reduce the dimensionality of the dataset while preserving or even enhancing model performance. Feature selection methods can be either manual, where domain expertise guides the selection process, or automatic, where algorithms are employed to identify the most significant features. Effective feature selection serves to simplify model complexity, mitigate overfitting, and enhance model interpretability. In this context, out of the 15 features available in the dataset, 3 features have been selected for analysis (as shown in Table III): Temperature, Relative Humidity and Oxygen level content of that particular area. Other features such as Day, Month, Year, Wind Speed, Rainfall, Fibe Fuel Moisture Code, Duff Moisture Code, Drought Code, Initial Spread Index, Buildup Index and Fire Weather Index were excluded or not selected due to various reasons such as redundancy, lack of significant correlation with Forest fire prediction, or potential multicollinearity issues.

| Full Name | Symbol |
|---|---|
| Temperature | temp |
| Relative Humidity | rh |
| Oxygen Level | oxy |

TABLE II: Selected Features

*1) Temperature:* Temperature is a critical factor in forest fire prediction, significantly influencing the likelihood and behavior of fires. High temperatures contribute to the drying of vegetation, reducing its moisture content. Dry vegetation, such as leaves, grass, and branches, is more combustible and can ignite more easily compared to moist vegetation. This increased flammability directly correlates with a higher risk of fire, as dry plant material provides ample fuel for the fire to spread rapidly once ignited. Elevated temperatures also lead to increased rates of evaporation from soil and vegetation, which reduces overall humidity levels. This creates drier conditions that are conducive to fire ignition and spread. Low humidity combined with high temperatures results in an environment where fires can start and propagate quickly. The reduced moisture content in the air and vegetation means that any spark or flame can potentially lead to a significant fire event.

Once a fire has ignited, higher temperatures can enhance its behavior, making it more intense and difficult to control. Fires burn more vigorously in hot conditions, consuming fuel more rapidly and generating more heat. This heat can further dry out surrounding vegetation, creating a feedback loop that exacerbates the fire's intensity and spread. The increased combustion rate in high temperatures means that the fire can grow quickly, posing greater challenges for firefighting efforts and increasing the potential for extensive damage.

*2) Relative Humidity:* Relative humidity (RH) is a crucial factor in forest fire prediction, significantly influencing the moisture content of both the air and vegetation, which in turn affects the likelihood of fire ignition and spread. High RH means more moisture in the air, which helps keep vegetation moist and less flammable. Conversely, low RH results in drier air, which can desiccate vegetation, making it more susceptible to ignition. Dry vegetation is a critical factor in forest fire risk, as it serves as fuel that can ignite easily and burn rapidly.

Low relative humidity levels create favorable conditions for fire ignition. When RH is low, the air's moisture-holding capacity decreases, and vegetation loses moisture quickly through evaporation. This desiccation lowers the ignition threshold of plant materials, meaning they can catch fire from smaller sparks or lower energy sources. Therefore, periods of low RH are often associated with higher fire danger, as the dry conditions make it easier for fires to start and spread. Relative humidity also influences the behavior of an ongoing fire. Low RH can lead to more intense and faster-spreading fires because the drier air and vegetation allow fires to consume fuel more rapidly.

The interaction between temperature and relative humidity is critical in fire prediction. High temperatures often coincide with low RH, compounding the risk factors for fire ignition and propagation. As temperature rises, the air can hold more moisture, but if the actual moisture content does not increase, the relative humidity drops, creating a more flammable environment. This combination of high temperature and low RH is particularly dangerous for fire risk, as it sets the stage for rapid ignition and aggressive fire behavior.

*3) Oxygen Levels:* Oxygen levels play a critical role in forest fire prediction as they directly influence the combustion process. The availability of oxygen is essential for sustaining and propagating fires. High oxygen levels can enhance the combustion process, making it easier for fires to ignite and spread. In contrast, low oxygen levels can inhibit combustion, making it more difficult for fires to sustain themselves. Therefore, monitoring oxygen levels can provide crucial insights into the potential for fire ignition and behavior.

The intensity of a fire is closely related to the concentration of oxygen in the surrounding air. Higher oxygen levels can lead to more vigorous and intense fires, as the increased availability of oxygen accelerates the combustion rate. This results in higher temperatures and more rapid consumption of fuel. Conversely, lower oxygen levels can reduce fire intensity, slowing the combustion process and making fires easier to control. In environments with high oxygen concentrations, fires can spread more rapidly as the increased oxygen supply fuels the combustion of adjacent materials. This can lead to larger and more destructive fires. On the other hand, areas with lower oxygen levels may experience slower fire spread, as the limited oxygen availability constrains the fire's ability to consume new fuel sources.

Moreover, oxygen levels interact with other environmental factors such as temperature and humidity to influence fire behavior. For instance, in hot and dry conditions, high oxygen levels can exacerbate the fire risk by promoting more efficient combustion of dry vegetation. Similarly, in environments where humidity is low, the presence of high oxygen levels can further enhance fire ignition and spread. Understanding these interactions is crucial for accurate fire prediction and management.

## D. "Data Splitting"

Data splitting is a critical step in the model development process, especially in supervised learning tasks. It involves dividing the available dataset into separate subsets for training, validation, and testing. The training set is used to train the model, the validation set is used to tune hyperparameters and evaluate model performance during training, and the testing set is used to assess the final performance of the trained model on unseen data. Proper data splitting helps prevent overfitting and provides an unbiased estimate of the model's performance on new data.

## E. "Model Selection"

Model selection is the process of choosing the most appropriate algorithm and model architecture for a specific task or dataset. It involves evaluating and comparing different models based on their performance metrics, such as accuracy, precision, recall, and F1-score. Model selection considers factors such as model complexity, computational efficiency, interpretability, and scalability. Techniques for model selection include cross-validation, grid search, and model evaluation on holdout datasets. The goal of model selection is to develop predictive models that generalize well to new data and effectively solve the problem at hand.

*1) Logistic Regression:*
**Step 1: Data Splitting:** In our Forest Fire Prediction model, the Logistic Regression algorithm begins by identifying the most relevant features to predict the likelihood of a fire. The process involves selecting features such as temperature, humidity, oxygen level, and other relevant environmental parameters.

The dataset is split into training and testing sets to evaluate the model's performance. Typically, an 80/20 split is used, where 80% of the data is used for training and 20% for testing.

**Logistic Regression Equation :**

$$\hat{p} = \frac{1}{1 + e^{-(\beta_0 + \beta_1 x_1 + \beta_2 x_2 + \cdots + \beta_n x_n)}}$$

where:
- $\hat{p}$ is the predicted probability of a fire occurring.
- $\beta_0$ is the intercept.
- $\beta_1, \beta_2, \ldots, \beta_n$ are the coefficients for each feature.
- $x_1, x_2, \ldots, x_n$ are the feature values.

**Step 2 : Training the model :** The Logistic Regression model is trained using the training dataset to learn the optimal values of the coefficients ($\beta$s). This is typically done using Maximum Likelihood Estimation (MLE), which maximizes the likelihood that the observed data occurred given the parameter values.

**Step 3 : Evaluation Metrics :** After training the model, the model's performance is evaluated using the following metrics:

**1. Accuracy :** Accuracy is the ratio of correctly predicted instances to the total instances, defined as:

$$\text{Accuracy} = \frac{TP + TN}{TP + TN + FP + FN}$$

where:
- $TP$ is the number of true positives.
- $TN$ is the number of true negatives.
- $FP$ is the number of false positives.
- $FN$ is the number of false negatives.

**Interpretation :** Higher accuracy indicates better model performance.

**2. Precision:** Precision in the context of classification models refers to the accuracy of positive predictions made by the model. It measures the proportion of correctly predicted positive cases (True Positives, TP) out of all instances that the model predicted as positive (True Positives + False Positives, TP + FP).

$$\text{Precision} = \frac{TP}{TP + FP}$$

where:
- $TP$ is the number of true positives.
- $FP$ is the number of false positives.

**Interpretation :** High precision indicates a low false positive rate.

**3. Recall (Sensitivity):** Recall, also known as Sensitivity or True Positive Rate (TPR), measures the ability of a classification model to correctly identify all positive instances, out of all instances that actually belong to the positive class. In other words, it calculates the proportion of true positives (correctly predicted positive instances) to all instances that are actually positive.

$$\text{Recall} = \frac{TP}{TP + FN}$$

**Interpretation :** High recall indicates a low false negative rate.

**4. F1 Score :** The F1 Score is a metric that combines both Precision and Recall into a single value, providing a balance between them. It is particularly useful in scenarios where the class distribution is imbalanced, meaning one class (e.g., positive or negative) is much more prevalent than the other.

$$\text{F1 Score} = \frac{2 \times \text{Precision} \times \text{Recall}}{\text{Precision} + \text{Recall}}$$

where:
- Precision is the number of true positives divided by the sum of true positives and false positives.
- Recall is the number of true positives divided by the sum of true positives and false negatives.

**Interpretation :** A higher F1 Score indicates a better balance between precision and recall.

**5. Receiver Operating Characteristic (ROC) Curve and Area Under the Curve (AUC):** The Receiver Operating Characteristic (ROC) curve is a graphical plot that illustrates the performance of a binary classification model across different threshold settings. It displays the true positive rate (TPR), also known as recall or sensitivity, on the y-axis and the false positive rate (FPR) on the x-axis.

$$AUC = \int_0^1 ROC(t)\,dt$$

where $ROC(t)$ represents the ROC curve.

**Interpretation :** A higher AUC indicates better model performance in distinguishing between positive and negative classes.

*2) Random Forest Classifier:*

**Step 1: Bootstrapping:** In our Forest Fire Prediction model, the process begins with bootstrapping, where we randomly sample the dataset with replacement to create multiple bootstrap samples. This means that each bootstrap sample may contain duplicate instances from the original dataset. By generating these bootstrap samples, we introduce variability into the training process, allowing decision trees to capture different aspects of the data.

**Bootstrapping (for creating bootstrap samples):**
- Randomly sample $n$ instances from the dataset with replacement to form a bootstrap sample.

**Step 2: Building Trees:** For each bootstrap sample, we build a decision tree using the Decision Tree Classifier algorithm. At each node of the tree, a subset of features is randomly selected for consideration when deciding how to split the data. This randomness helps prevent overfitting and promotes diversity among the trees.

**Step 3: Aggregation:** In a Random Forest model, multiple decision trees are constructed using different bootstrap samples from the original dataset. Each of these decision trees operates independently and provides its own prediction based on the features of the input data. For regression tasks, the final prediction is obtained by averaging the predictions of all these individual decision trees. For classification tasks like Forest fire prediction, we use majority voting to determine the final class label. This means that each tree's prediction is considered, and the class that receives the most votes across all trees is chosen as the final prediction This aggregation approach helps improve the robustness and generalization of the model.

**Aggregation (for combining predictions of decision trees):**
- For classification tasks: Majority voting is used to determine the final class label.
- For regression tasks: Averaging is used to calculate the final prediction.

By employing this ensemble method, our Forest Fire Prediction model benefits from the collective wisdom of multiple decision trees, resulting in more accurate and reliable predictions. The randomness introduced during bootstrapping and feature selection, combined with the aggregation of predictions, enhances the model's ability to effectively predict fire risk based on environmental data.

*3) Support Vector Machine:*

**Step 1: Margin Maximization:** In our Forest Fire Prediction model, the Support Vector Machine (SVM) algorithm starts by identifying the hyperplane that maximizes the margin between classes in the feature space. The margin represents the distance between the hyperplane and the nearest data points, known as support vectors, from each class. By maximizing the margin, SVM aims to find the optimal decision boundary that best separates high fire risk and low fire risk instances in the feature space.

**Step 2: Kernel Trick:** If the forest fire data is not linearly separable in its original feature space, the SVM algorithm employs a kernel trick to map the data to a higher-dimensional space where it may become linearly separable. This is achieved using kernel functions such as polynomial or radial basis function (RBF) kernels. By transforming the data into a higher-dimensional space, SVM seeks to find a separating hyperplane that effectively separates high fire risk and low fire risk instances.

**Step 3: Optimization:** Once the data is mapped to a suitable space, SVM solves an optimization problem to determine the optimal hyperplane parameters (weights and bias) that maximize the margin between classes while minimizing classification errors. By finding the optimal hyperplane, SVM ensures robust and accurate classification of forest fire risk based on environmental features.

**Decision Function:** The decision function in SVM, denoted as $f(x)$, determines the predicted class label for a given input feature vector $x$. It calculates the sign of the dot product between the weight vector $w$ and the input feature vector $x$, plus the bias term $b$. The decision function assigns high fire risk or low fire risk labels based on the calculated sign.

$$f(x) = \text{sign}(w \cdot x + b)$$

Where:
- $f(x)$ is the decision function.
- $w$ is the weight vector.
- $x$ is the input feature vector.
- $b$ is the bias term.

**Margin Calculation:** In our forest fire prediction model, the margin is computed as the inverse of twice the Euclidean norm of the weight vector $||w||$. This calculation quantifies the distance between the decision boundary (hyperplane) and the support vectors, providing a measure of separability between high and low fire risk classes.

$$\text{Margin} = \frac{2}{||w||}$$

Where:
- $||w||$ is the Euclidean norm of the weight vector.

**Hinge Loss (for binary classification):** The hinge loss function is utilized in SVM for binary classification tasks, such as classifying fire risk instances. It measures the classification error for each sample based on the decision function output. The hinge loss penalizes misclassifications by computing the maximum of zero and $1 - y_i \cdot (w \cdot x_i + b)$, where $y_i$ represents the true class label of sample $i$ and $(w \cdot x_i + b)$ is the decision function output. This loss function guides the optimization process towards minimizing classification errors while maximizing the margin.

$$\text{Hinge Loss} = \max(0, 1 - y_i \cdot (w \cdot x_i + b))$$

Where:
- $y_i$ is the true class label of sample $i$.
- $(w \cdot x_i + b)$ is the decision function output.

### F. "Classification"

In this section, we present the methodology used for predicting forest fire risk based on environmental data attributes, including temperature, humidity, oxygen levels, and other relevant factors.

To predict forest fire risk, we developed a classification model using machine learning algorithms trained on the pre-processed dataset, which includes various meteorological and environmental features.

The model was trained using a supervised learning approach, where the target variable was the level of fire risk (e.g., high risk or low risk). We employed various classification algorithms such as Logistic Regression, Random Forest Classifier, and Support Vector Machine (SVM) Classifier.

*1) Environmental Data Attributes:*
- **Temperature :** The ambient temperature, which plays a critical role in determining the likelihood of a fire starting and spreading.
- **Humidity :** The level of moisture in the air, which affects the dryness of vegetation and the ease with which it can ignite.
- **Oxygen Level :** The oxygen content in the atmosphere, which influences the spread rate of the fire and its direction.

By employing these machine learning classification algorithms, our forest fire prediction model effectively classifies and predicts the risk of forest fires based on a variety of environmental data attributes. This enables early detection and preventative measures to be taken, potentially reducing the impact of forest fires.

## IV. HARDWARE COMPONENTS

In this study, we have utilized a variety of hardware components to achieve real-time fire detection, including a DHT11 sensor, an IR flame sensor, an Arduino UNO module, a GSM module, and a buzzer. Each component plays a crucial role in ensuring accurate detection and timely alerts.

- The DHT11 sensor is employed for monitoring environmental temperature and humidity.

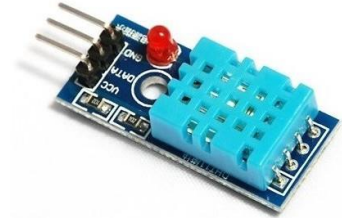

Fig. 1: DHT11 Sensor

- The IR flame sensor is a critical component used for detecting fire. It can detect infrared light with a wavelength between 760 nm and 1100 nm, which is typical of flame signatures.

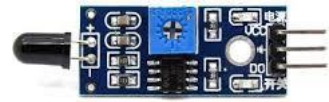

Fig. 2: IR Flame Sensor

- The Arduino UNO module collects data from the DHT11 and IR flame sensors, processes the information, and triggers the GSM module to send alert messages and activate the buzzer upon detecting a fire.

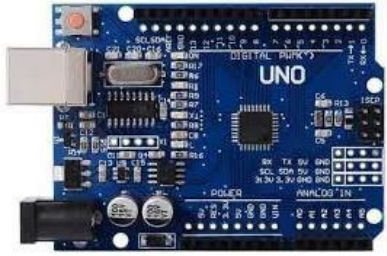

Fig. 3: Arduino UNO module

- The GSM module is integrated in the system for real time alerting and sending alert messages upon fire detection.

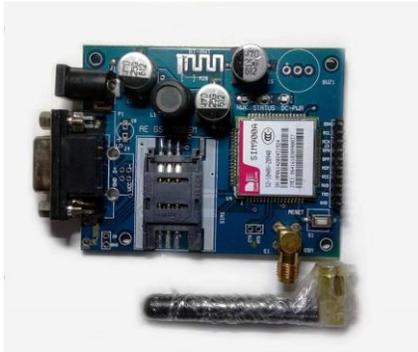

Fig. 4: SIM Toolkit GSM module

- The buzzer is activated immediately upon the detection of smoke or fire, providing a local alert to individuals in the vicinity.

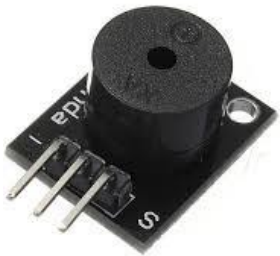

Fig. 5: Buzzer

- In the final circuit connection, the DHT11 sensor and IR flame sensor are connected to the digital input pins of the Arduino Uno, allowing it to read temperature, humidity, and flame presence data. The GSM module is interfaced with the Arduino via the UART (TX and RX) pins for sending SMS alerts. The buzzer is connected to a digital output pin to be activated by the Arduino when a fire is detected.

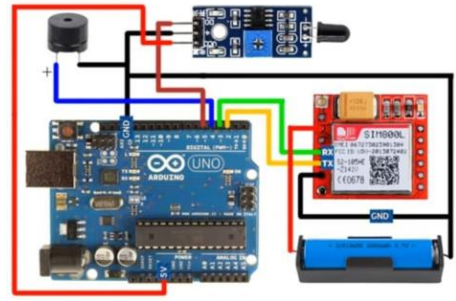

Fig. 6: Circuit Diagram of the connection

## V. RESULT AND DISCUSSION

In this section, we outline our approach to predicting forest fires based on environmental data attributes, notably incorporating temperature, humidity, and oxygen level. Our workflow encompasses comprehensive data preprocessing steps, including the encoding of categorical variables and scaling of numerical features, to ensure optimal model performance. Subsequently, leveraging these preprocessed features, we employ a suite of machine learning algorithms tailored for both regression and classification tasks. Specifically, we train three distinct models: Logistic Regression, Random Forest Classifier, and Support Vector Machine (SVM) Classifier. For classification tasks, we utilize the area under the Receiver Operating Characteristic (ROC) curve (AUC), in addition to traditional metrics such as accuracy, precision, and recall. These metrics provide a comprehensive evaluation of our models' ability to discriminate between different classes and their overall performance in classifying forest fire risk.

- **Temperature:** Ambient temperature in degrees Celsius.
- **Humidity:** Relative humidity as a percentage.
- **Oxygen Level:** Concentration of oxygen in the air as a percentage.

| Full Name | Symbol |
|---|---|
| Temperature | temp |
| Relative Humidity | rh |
| Oxygen Level | oxy |

TABLE III: Selected Features

Table IV summarizes the performance metrics of the classification models:

| Model | Accuracy | Precision | Recall | F1 Score |
|---|---|---|---|---|
| Logistic Regression | 89.00 | 0.66 | 0.93 | 0.88 |
| Random Forest | 93.00 | 0.91 | 0.83 | 0.89 |
| Support Vector | 93.00 | 0.92 | 1.00 | 0.93 |

TABLE IV: Model Evaluation Metrics

Based on the accuracy scores obtained, the Support Vector Classifier outperforms the other models, achieving an accuracy of 93%. Therefore, we choose the Support Vector Machine for further analysis and interpretation.

### A. Confusion Matrix of Support Vector Machine

The confusion matrix provides a detailed summary of the classifier's performance by showing the true positive, true negative, false positive, and false negative predictions. Each cell in the matrix represents the count of instances classified accordingly. The heatmap visualization adds clarity by highlighting the distribution of these counts. The rows of the matrix correspond to the actual classes, while the columns represent the predicted classes. Annotated values within the cells indicate the number of instances falling into each category.

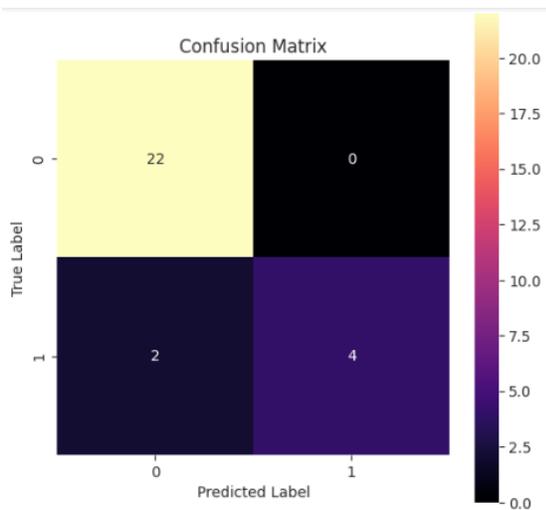

Fig. 7: Confusion Matrix of Support Vector Machine

### B. Discussion

In the confusion matrix, each cell represents a specific scenario:
- **True Positives (TP)**: These are the instances where the model correctly identifies areas with a high risk of forest fires.
- **True Negatives (TN)**: These are the instances where the model correctly identifies areas with a low risk of forest fires.
- **False Positives (FP)**: These are the instances where the model incorrectly predicts a high risk of forest fire when the area is actually at a low risk.
- **False Negatives (FN)**: These are the instances where the model incorrectly predicts a low risk of forest fire when the area is actually at a high risk.

Here's what we can infer from the confusion matrix:
- **Accuracy**: This metric tells us how often the model correctly predicts the risk level of forest fires overall. It's the percentage of correct predictions out of all predictions made.
- **Precision**: This metric measures how many of the predicted high-risk areas for forest fires are actually at high risk. It's important because it tells us the proportion of correct high-risk predictions.
- **Recall (Sensitivity)**: This metric tells us how many of the actual high-risk areas for forest fires the model captured. It's important because it shows the model's ability to detect high-risk areas accurately.
- **F1 Score**: This metric combines precision and recall. It's useful for finding a balance between these two metrics, indicating an overall performance measure that considers both false positives and false negatives.

By analyzing these metrics, we can assess how well our forest fire prediction model, such as a Support Vector Classifier, performs in identifying areas at risk of forest fires. It provides insights into where the model excels and where it requires improvement in accurately predicting the risk of forest fires, thereby aiding in forest management and fire prevention efforts.

### C. Integration of the Hardware Components

It was observed that the final circuit connection, when all hardware components were integrated correctly, performed as expected. The DHT11 sensor and IR flame sensor successfully detected temperature, humidity, and the presence of flames. Upon fire detection, the Arduino Uno promptly activated the buzzer, providing an immediate local alert. Simultaneously, the GSM module sent an alert SMS to authorized personnel, confirming the system's capability for real-time fire detection and notification. This demonstrates the reliability and effectiveness of our integrated fire detection system.

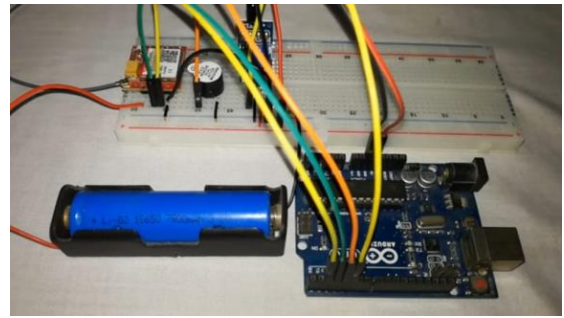

Fig. 8: Connection diagram of the hardware components

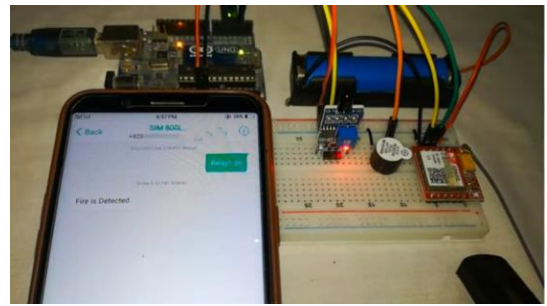

Fig. 9: SMS alert sent to the user

*D. Website Interface and Output:*

We designed a user-friendly web interface tailored for Forest fire prediction, aimed at providing valuable insights into the likelihood of forest fires based on relevant environmental data. Once users submit the required environmental data which includes temperature, humidity and oxygen content, our prediction model processes the information and generates output regarding the probability of forest fire occurrence in the specified area. This output provides valuable insights into the risk level of forest fires, empowering users to make informed decisions and take proactive measures for forest management and fire prevention. Screenshots of the website interface and sample output are provided in Figure 10.

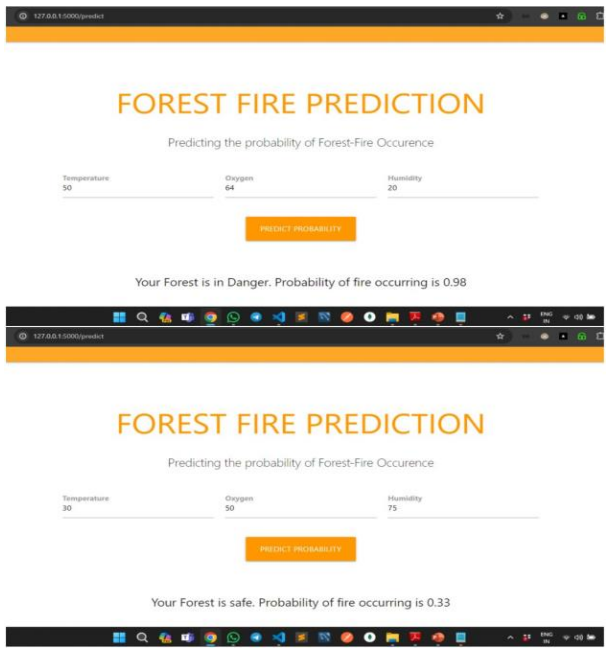

Fig. 10: Webpage: Visualizations and Predictive Results for Forest Fire Prediction

## VI. CONCLUSION

We evaluated the performance of three prominent classifiers: Logistic Regression, Random Forest, and Support Vector Machine (SVM), to predict forest fire occurrences based on the sensor data. Our findings indicate promising results with high accuracy scores of 89%, 92.67%, and 93% for Logistic Regression, Random Forest, and SVM, respectively

In conclusion, our project represents a significant stride forward in both forest fire prediction methodologies and the broader realm of data science applications. By employing a structured and iterative approach, we have not only developed a functional system for forest fire detection but also deepened our understanding of the intricate relationship between environmental variables and fire behavior. Through the utilization of machine learning techniques, we have gleaned invaluable insights into the predictive patterns inherent in temperature, humidity, and other environmental factors, shedding light on the complex dynamics of forest ecosystems under fire risk conditions.

Our study focused on developing a predictive model for forest fire occurrence using machine learning algorithms. We utilized a comprehensive dataset comprising environmental variables such as temperature, humidity, oxygen concentration and other topographical features. Through rigorous analysis and experimentation, we assessed the performance of three prominent classifiers: Logistic Regression, Random Forest, and Support Vector Machine (SVM)..

Upon comparing the performance of these classifiers, the Support Vector model emerged as the most promising one, exhibiting superior accuracy, precision, recall, and F1 score. The Support Vector Machine algorithm is known for its robustness and ability to handle complex datasets, making it suitable for our Forest Fire prediction task.

Furthermore, we conducted an analysis of the confusion matrix for the Support Vector classifier, which provided insights into its predictive performance. The confusion matrix revealed the model's ability to correctly classify fire and not-fire cases, along with any misclassifications or errors.

Overall, our findings underscore the promise of machine learning methodologies, with the Support Vector Machine algorithm showing notable potential in accurately predicting forest fire occurrence based on environmental variables. However, it's imperative to acknowledge the need for further validation and refinement of the predictive models before considering their deployment in operational settings. Future research endeavors could concentrate on augmenting the dataset with additional relevant features, fine-tuning model hyperparameters, and conducting robust validation assessments to bolster the models' predictive capabilities and overall reliability in real-world forest fire prediction scenarios.

The predictions generated by the forest fire prediction model can empower forest management authorities and firefighting teams in early detection and proactive management of forest fires, thereby mitigating their destructive impact on ecosystems and human settlements. By providing timely alerts and actionable insights, the model facilitates swift response strategies, enabling authorities to allocate resources effectively, implement preventive measures, and coordinate evacuation efforts when necessary. Ultimately, these proactive measures contribute to minimizing the spread of fires, reducing property damage, protecting biodiversity, and safeguarding human lives and infrastructure.